\documentclass[
fleqn]{elsart}
\usepackage{amsmath,amssymb}

\def\be{\begin{equation}}
\def\ee{\end{equation}}
\def\bea{\begin{eqnarray}}
\def\eea{\end{eqnarray}}
\def\ba{\begin{array}}
\def\ea{\end{array}}
\def\nn{\nonumber}

\def\sda{\sin^2\theta}

\def\Dp{\Delta^{\prime}}
\def\Dpp{\Delta^{\prime\prime}}
\def\trm{\textrm{Im}}

\begin{document}
\begin{flushright}
hep-th/0511123~v5
\end{flushright}

\begin{frontmatter}

\title{Hawking radiation of charged particles as tunneling from Reissner-Nordstr\"{o}m-de
Sitter black holes with a global monopole}
\author{Qing-Quan Jiang}
\address{College of Physical Science and Technology, Central China Normal University,
Wuhan, Hubei 430079, People's Republic of China \\
Institute of Theoretical Physics, China West Normal University,
Nanchong, Sichuan 637002, People's Republic of China}
\author{Shuang-Qing Wu}
\ead{sqwu@phy.ccnu.edu.cn, Corresponding author}
\address{College of Physical Science and Technology, Central China Normal University,
Wuhan, Hubei 430079, People's Republic of China}
\date{today}

\begin{abstract}
Applying Parikh's semi-classical tunneling method, we consider Hawking radiation of the charged
massive particles as a tunneling process from the Reissner-Nordstr\"{o}m-de Sitter black hole with
a global monopole. The result shows that the tunneling rate is related to the change of Bekenstein-Hawking
entropy and the radiant spectrum is not a pure thermal one, but is consistent with an underlying
unitary theory.
\end{abstract}

\end{frontmatter}

\newpage

\section{Introduction}

In 1974, Hawking \cite{Hawking} proved that the black hole can emit particles from its event
horizon with a temperature proportional to its surface gravity, and the radiant spectrum is a
pure thermal one, which implies the loss of information of black hole after it has evaporated
away and disappeared completely \cite{ILP}. Though a complete resolution of the information loss
paradox must be in the framework of quantum gravity and/or the unitary theory of string/M-theory,
Hawking argued that the information could come out if the outgoing radiation were not exactly
thermal but had subtle corrections.

Recently, Parikh and Wilczek \cite{PW} put forward a semi-classical tunnelling method to investigate
Hawking radiation of the static Schwarzschild and Reissner-Nordstr\"{o}m black holes, they found that
the radiant spectrum of the black hole is not a pure thermal one and the derived tunneling rate is
related to the change of Bekenstein-Hawking entropy. In their methodology, Hawking radiation is
treated as a tunneling process with the tunneling potential barrier produced by the outgoing particle
itself. The key trick to calculate the tunneling rate is to find a coordinate system well-behaved
at the event horizon. However, this method is currently limited to discuss the tunneling rate of
the uncharged massless particles only \cite{PW,TRA,ZZY}. For black holes with a charge, the
emitted outgoing particles can be charged also, not only should the energy conservation but also
the charge conservation be considered \cite{KWZ}.

On the other hand, researches on the charged black hole with a positive cosmological constant and
with a global monopole become important due to the following reasons: (1) The recent observed
accelerating expansion of our universe indicates the cosmological constant might be a positive
one \cite{AEU}; (2) Conjecture about de Sitter/conformal field theory (CFT) correspondence \cite{dSCFT};
(3) There might exist topological defects in the early universe \cite{TSY}; etc.

Combined with the reasons mentioned above, in this Letter we extend the Parikh's method to investigate
the Hawking radiation of the charged particle via tunneling from the Reissner-Nordstr\"{o}m-de Sitter
black hole with a global monopole whose Arnowitt-Deser-Misner (ADM) mass is $(1 -8\pi\eta^2)$ times
than that of mass parameter. Our result shows that the emission rate of the charged particle is
connected with the Bekenstein-Hawking entropy, and the corrected radiant spectrum is not a pure
thermal one, but is consistent with an underlying unitary theory.

Our Letter is outlined as follows: In Section \ref{GPRG}, we introduce the generalized Painlev\'{e} coordinate
transformation and present the radial geodesic equation of charged particles. In Sections \ref{TREH} and
\ref{TRCH}, we investigate Hawking radiation as tunneling from the event horizon and the cosmological
horizon, and compute the tunneling rate from these two horizons, respectively. Finally we give some
discussions about our results.

\section{Generalized Painlev\'{e} coordinate transformation and
 the radial geodesics of charged particles}
\label{GPRG}

The line element of a Reissner-Nordstr\"{o}m-de Sitter black hole with a global monopole is \cite{GS}
\be
ds^2 = -\Delta dt_R^2 +\Delta^{-1}dr^2 +(1 -8\pi\eta^2)r^2(d\theta^2 +\sda d\phi^2) \, ,
\label{RNdSM}
\ee
where $\Delta = 1 -2M/r +Q^2/r^2 -(\Lambda/3)r^2$, $\eta$ is a symmetry breaking constant related to the
global monopole, $M$ is the mass parameter, $Q$ is the charge of the black hole, $\Lambda$ is a positive
cosmological constant, and $t_R$ is the coordinate time for the black hole. In general, the black hole
has an inner horizon (IH), an event horizon (EH) and a cosmological horizon (CH), all of them satisfying
the horizon equation $\Delta = 0$. In this Letter we shall consider the most general case where neither of
these horizons coincides with the other one.

To remove the coordinate singularity in the metric (\ref{RNdSM}), we introduce a generalized Painlev\'{e}
coordinate transformation
\be
dt_R = dt \mp \frac{\sqrt{1 -\Delta}}{\Delta}dr \, ,
\ee
and obtain the Painlev\'{e}-like line element of the Reissner-Nordstr\"{o}m-de Sitter black hole with a
global monopole as follows
\be
ds^2 = -\Delta dt^2 \pm 2\sqrt{1 -\Delta}dtdr +dr^2 +(1 -8\pi\eta^2)r^2(d\theta^2 +\sda d\phi^2) \, ,
\label{metric}
\ee
where a plus (minus) sign denotes the space-time line element of the charged massive outgoing (ingoing)
particles at the EH (CH). In Eq. (\ref{metric}), the Painlev\'{e}-like coordinate system has many
attractive features. First, the metric is well behaved at the EH and CH; Secondly, it satisfies Landau's
condition of the coordinate clock synchronization; Thirdly, the new form of the line element is stationary,
but not static. These characters are useful to investigate the tunneling radiation of the charged massive
particles across the horizons.

It should be pointed out that unlike the asymptotically flat case, the Painlev\'{e}-like coordinate for the
asymptotically non-flat space-time is not unique. In fact, there is another form for the metric (\ref{RNdSM})
\bea
ds^2 &=& -\Delta dt^2 \pm 2\sqrt{1 -\Delta/(1 -\Lambda r^2/3)}dtdr +(1 -\Lambda r^2/3)^{-1}dr^2 \nn \\
&&\qquad +(1 -8\pi\eta^2)r^2(d\theta^2 +\sda d\phi^2) \, , \nn
\eea
which approaches to the de Sitter space in the vacuum case where $\eta = 0$.

Now, let us work with the metric in the new form (\ref{metric}) and obtain the radial geodesics of the
charged massive particles, which is different from that of the uncharged massless particles that follow
the radial null geodesics
\be
\dot{r} = \frac{dr}{dt} = \pm 1 \mp \sqrt{1 -\Delta} \, .
\ee
According to de Broglie's hypothesis, from the definition of the phase velocity $v_p$ and the group velocity
$v_g$, we have
\be
v_p = \frac{1}{2}v_g \, .
\ee
Since the tunneling process is an instantaneous effect, the metric in the line element (\ref{metric})
satisfies Landau's condition of the coordinate clock synchronization, the coordinate time difference of
two events, which take place simultaneously in different places, is
\be
dt = -\frac{g_{tr}}{g_{tt}}dr_c \, , \qquad\qquad (d\theta = d\phi = 0) \, ,
\ee
where $dr_c$ is the location of the tunneling particle. So the group velocity can be expressed as
\be
v_g = \frac{dr_c}{dt} = -\frac{g_{tt}}{g_{tr}}
= \pm \frac{r^2 -2Mr +Q^2 -(\Lambda/3)r^4}{\sqrt{2Mr^3 -Q^2r^2 +(\Lambda/3)r^6}} \, ,
\ee
therefore the phase velocity (the radial geodesics) is
\be
\dot{r} = v_p = -\frac{g_{tt}}{2g_{tr}}
= \pm \frac{r^2 -2Mr +Q^2 -(\Lambda/3)r^4}{2\sqrt{2Mr^3 -Q^2r^2 +(\Lambda/3)r^6}} \, ,
\ee
where $+ (-)$ sign denotes the phase velocity of the charged particles tunneling across the EH (CH).
During the process of a charged massive particle tunneling across the potential barrier, the self-interaction
effect of the electro-magnetic field on the emitted particles should not be ignored, and the temporal
component of electro-magnetic potential is
\be
A_t = \pm \frac{Q}{r} \, .
\ee

In the remaining two sections, we shall discuss Hawking radiation from the event horizon and the cosmological
horizon, respectively, and calculate the tunneling rate from each horizon. Since the overall picture of
tunneling radiation for the metric is very involved, to simplify the discussion we will consider the outgoing
radiation from the EH, and ignore the incoming radiation from the CH, for the moment when we deal with the
black hole event horizon. While dealing with the CH case, we shall only consider the incoming radiation from
the CH and ignore the outgoing radiation from the EH.

\section{Tunneling rate of charged particles at the EH}
\label{TREH}

According to the energy conservation and the charge conservation, one can assume that the total ADM mass and charge
of the hole-particle system are held fixed whereas the mass and the charge of the hole are allowed to fluctuate,
the black hole mass and charge will become $M -\omega$, $Q -q$ when a particle with energy $\omega$ and charge
$q$ has evaporated from the EH. Considering the charged particle tunnels out from the EH along the radial direction,
we can get the new line element of the black hole in the EH case
\be
ds^2 = -\Dp dt^2 +2\sqrt{1 -\Dp}dtdr +dr^2 +(1 -8\pi\eta^2)r^2(d\theta^2 +\sda d\phi^2) \, ,
\ee
where $\Dp = 1 -2(M -\omega)/r +(Q -q)^2/r^2 -(\Lambda/3)r^2$. Accordingly the radial geodesics of the charged
massive particles tunneling out from the EH is
\be
\dot{r} = \frac{r^2 -2(M -\omega)r +(Q -q)^2 -(\Lambda/3)r^4}{2\sqrt{2(M -\omega)r^3
-(Q -q)^2r^2 +(\Lambda/3)r^6}} \, ,
\label{ME1}
\ee
and the non-zero component of electro-magnetic potential becomes
\be
A_t = \frac{Q -q}{r} \, .
\ee

When the charged particle tunnels out, the effect of the electro-magnetic field should be taken into account.
So the matter-gravity system consists of the black hole and the electro-magnetic field outside the hole. As the
Lagrangian function of the electro-magnetic field corresponding to the generalized coordinates described by
$A_{\mu}$ is $-(1/4)F_{\mu\nu}F^{\mu\nu}$, we can find that the generalized coordinate $A_{\mu} = (A_t, 0, 0, 0)$
is an ignorable coordinate. In order to eliminate the degree of freedom corresponding to $A_{\mu}$, the imaginary
part of the action for the charged massive particle should be written as
\bea
\trm S &=& \trm\int_{t_i}^{t_f}\big(L -P_{A_t}\dot{A_t}\big)dt
= \trm\int_{r_{ie}}^{r_{fe}}\big(P_r\dot{r} -P_{A_t}\dot{A_t}\big)\frac{dr}{\dot{r}} \nn \\
&=& \trm\int\limits_{r_{ie}}^{r_{fe}}\Bigg[\int\limits_{(0, ~0)}^{(P_r, P_{A_t})}
\Big(\dot{r}~dP_r^{\prime} -\dot{A_t}~dP_{A_t}^{\prime}\Big)\Bigg]\frac{dr}{\dot{r}} \, ,
\label{IA1}
\eea
where $r_{ie}$ and $r_{fe}$ represent the locations of the EH before and after the particle with energy
$\omega$ and charge $q$ tunnels out. According to Hamilton's canonical equation of motion, we have
\bea
\dot{r} &=& \frac{dH}{dP_r}\Big|_{(r; A_t, P_{A_t})} \, , \qquad
dH|_{(r; A_t, P_{A_t})} = (1 -8\pi\eta^2)d\big(M -\omega\big) \, , \nn \\
\dot{A_t} &=& \frac{dH}{dP_{A_t}}\Big|_{(A_t; r, P_r)} \, , \qquad
dH|_{(A_t; r, P_r)} = (1 -8\pi\eta^2)\frac{Q -q}{r}d(Q -q) \, ,
\label{HCE1}
\eea
where $\omega$ and $q$ are the energy and the charge of the emitted particle. Because of the existence of
a global monopole in the black hole background, the total ADM mass and the total charge in the EH case are
$M_{\infty} = (1 -8\pi\eta^2)M$ \cite{MC} and $Q_{\infty} = (1 -8\pi\eta^2)Q$, respectively. [For the sake
of convenience, we take the mass and charge of the particle measured at infinity as $\omega_{\infty} =
(1 -8\pi\eta^2)\omega$ and $q_{\infty} = (1 -8\pi\eta^2)q$.] Eq. (\ref{HCE1}) represents the energy change
of the hole because of the loss of mass and charge when a particle tunnels out. Substituting Eqs. (\ref{ME1})
and (\ref{HCE1}) into Eq. (\ref{IA1}) and switching the order of integral, we obtain
\bea
\trm S &=& \trm \int\limits_{r_{ie}}^{r_{fe}}
\int\limits_{(1 -8\pi\eta^2)(M, ~Q)}^{(1 -8\pi\eta^2)(M -\omega, ~Q -q)}
\Big[dH|_{(r; A_t, P_{A_t})} -dH|_{(A_t; r, P_r)}\Big] \frac{dr}{\dot{r}} \nn \\
&=& \trm \int\limits_{(1 -8\pi\eta^2)(M, ~Q)}^{(1 -8\pi\eta^2)(M -\omega, ~Q -q)}
\int\limits_{r_{ie}}^{r_{fe}} \frac{2\sqrt{2(M -\omega^{\prime})r^3 -(Q -q^{\prime})^2r^2
+(\Lambda/3)r^6}}{r^2 -2(M -\omega^{\prime})r +(Q -q^{\prime})^2 -(\Lambda/3)r^4} \nn \\
&&\qquad \times (1 -8\pi\eta^2)\Big[d(M -\omega^{\prime})
-\frac{Q -q^{\prime}}{r}d(Q -q^{\prime})\Big] dr \, .
\label{IE1}
\eea

Since $1 -2(M -\omega^{\prime})/r +(Q -q^{\prime})^2/r^2 -(\Lambda/3)r^2 = 0$ satisfies the horizon equation
after the particle with energy $\omega^{\prime}$ and charge $q^{\prime}$ tunnels out, there exists a single
pole in Eq. (\ref{IE1}). Let us carry out the integral by deforming the contour around the pole so as to ensure
that the positive energy solutions decay in time, and get
\be
\trm S = -(1 -8\pi\eta^2)\trm\int_{r_{ie}}^{r_{fe}}(i\pi r)dr
= -\frac{\pi}{2}(1 -8\pi\eta^2)\big(r_{fe}^2 -r_{ie}^2\big) \, .
\ee
So the relationship between the tunneling rate and the imaginary part of the particle's action is
\be
\Gamma \sim e^{-2\trm S} = e^{\pi(1 -8\pi\eta^2)(r_{fe}^2 -r_{ie}^2)}
= e^{(A_{fe} -A_{ie})/4} = e^{\Delta S_{EH}} \, ,
\label{TA1}
\ee
where $A_{ie}$ and $A_{fe}$ denote the event horizon area before and after the charged particle tunnels
out, and $\Delta S_{EH}$ is the change of Bekenstein-Hawking entropy. From Eq. (\ref{TA1}), we find that
the tunneling rate at the EH is related to the Bekenstein-Hawking entropy, and is consistent with an
underlying unitary theory.

\section{Tunneling rate of charged particles at the CH}
\label{TRCH}

In this section, we will discuss the Hawking radiation of the charged particle via tunneling at the CH.
Different from the particle's tunneling behavior in the EH case discussed in the last section, the particle
is found to tunnel into the CH. So when the particle with energy $\omega$ and charge $q$ tunnels into
the CH, we can get the new line element as follows
\be
ds^2 = -\Dpp dt^2 -2\sqrt{1 -\Dpp}dtdr +dr^2 +(1 -8\pi\eta^2)r^2(d\theta^2 +\sda d\phi^2) \, ,
\ee
where $\Dpp = 1 -2(M +\omega)/r +(Q +q)^2/r^2 -(\Lambda/3)r^2$. Using the same method, the phase
velocity (the radial geodesics) of the charged particle tunneling into the CH can be expressed as
\be
\dot{r} = -\frac{r^2 -2(M +\omega)r +(Q +q)^2 -(\Lambda/3)r^4}{2\sqrt{2(M +\omega)r^3
-(Q +q)^2r^2 +(\Lambda/3)r^6}} \, ,
\label{ME2}
\ee
and the electro-magnetic potential becomes accordingly as
\be
A_t = -\frac{Q +q}{r} \, .
\ee

According to Hamilton's canonical equation of motion, when a particle with energy $\omega$ and charge $q$
is absorbed by the CH of the black hole, we can get
\bea
dH|_{(r; A_t, P_{A_t})} &=& -(1 -8\pi\eta^2)d\big(M +\omega\big) \, , \nn \\
dH|_{(A_t; r, P_r)} &=& -(1 -8\pi\eta^2)\frac{Q +q}{r}d(Q +q) \, ,
\label{HCE2}
\eea
where $\omega$ and $q$ are the energy and the charge of the absorbed particle. Due to the presence
of a global monopole in the black hole background, the total ADM mass and the total charge in the CH
case are $M_{\infty} = -(1 -8\pi\eta^2)M$ and $Q_{\infty} = -(1 -8\pi\eta^2)Q$, respectively. In the
same way, the imaginary part of the action for the charged massive particle incoming from the CH can
be expressed as
\bea
\trm S &=& \trm\int_{t_i}^{t_f}\big(L -P_{A_t}\dot{A_t}\big)dt
= \trm\int_{r_{ic}}^{r_{fc}}\big(P_r\dot{r} -P_{A_t}\dot{A_t}\big)\frac{dr}{\dot{r}} \nn \\
&=& \trm \int\limits_{-(1 -8\pi\eta^2)(M, ~Q)}^{-(1 -8\pi\eta^2)(M +\omega, ~Q +q)}
\int\limits_{r_{ic}}^{r_{fc}} \frac{2\sqrt{2(M +\omega^{\prime})r^3 -(Q +q^{\prime})^2r^2
+(\Lambda/3)r^6}}{r^2 -2(M +\omega^{\prime})r +(Q +q^{\prime})^2 -(\Lambda/3)r^4} \nn \\
&&\qquad \times (1 -8\pi\eta^2)\Big[d(M +\omega^{\prime})
-\frac{Q +q^{\prime}}{r}d(Q +q^{\prime})\Big] dr \, .
\label{IE2}
\eea
In Eq. (\ref{IE2}), $r_{ic}$ and $r_{fc}$ are the locations of the CH before and after the particle
with energy $\omega$ and charge $q$ is absorbed by the CH, and we find that $1 -2(M +\omega^{\prime})/r
+(Q +q^{\prime})^2/r^2 -(\Lambda/3)r^2 = 0$ is the horizon equation after the particle tunnels into the
CH, so there exists a single pole in Eq. (\ref{IE2}). Deforming the contour around the pole and carrying
out the integral, we have
\be
\trm S = -(1 -8\pi\eta^2)\trm\int_{r_{ic}}^{r_{fc}}(i\pi r)dr
= -\frac{\pi}{2}(1 -8\pi\eta^2)\big(r_{fc}^2 -r_{ic}^2\big) \, .
\ee
So the tunneling rate at the CH is
\be
\Gamma \sim e^{-2\trm S} = e^{\pi(1 -8\pi\eta^2)(r_{fc}^2 -r_{ic}^2)}
= e^{(A_{fc} -A_{ic})/4} = e^{\Delta S_{CH}} \, ,
\label{TA2}
\ee
where $A_{ic}$ and $A_{fc}$ are the cosmological horizon area before and after the charged massive
particle tunnels into the CH, and $\Delta S_{CH}$ is the change of the Bekenstein-Hawking entropy at
the CH. From Eq. (\ref{TA2}), we learn that tunneling rate at the CH of the Reissner-Nordstr\"{o}m-de
Sitter black hole with a global monopole is connected with Bekenstein-Hawking entropy.

\section{Summary and Discussions}

In summary, we find that when the charged massive particle tunnels across the event horizon (EH) and
the cosmological horizon (CH) of a Reissner-Nordstr\"{o}m-de Sitter black hole with a global monopole,
the radiant spectrum is not a pure thermal one, the tunneling rate is related to the change of
Bekenstein-Hawking entropy corresponding to each horizon, and is consistent with an underlying unitary
theory. So the Hawking radiation can be viewed as an ideal case only, it is possible for a not precisely
thermal radiation to carry out information during the radiation process of the black holes, and the
underlying unitary theory is reliable. The result obtained in this paper provides further evidence to
support the Parikh's tunneling picture, which might serve as a mechanism to deal with the information
loss paradox.

We would like to point out that a large class of previous results existed in the literature can be
enclosed as special case of ours obtained here. In particular, results obtained in Ref. \cite{PW,TRA}
can be recovered. For example, in the case where $\Lambda = 0$ and $\eta = 0$, the Reissner-Nordstr\"{o}m-de
Sitter black hole with a global monopole reduces to the Reissner-Nordstr\"{o}m black hole. Considering
an uncharged massless particle but with energy $\omega$ tunnels across the event horizon, we know that
$r_i = M +\sqrt{M^2 -Q^2}$ and $r_f = M -\omega +\sqrt{(M -\omega)^2 -Q^2}$ are the horizons of the
black hole before and after the emission of the particle. According to Eq. (\ref{TA1}), the tunneling
rate is
\be
\Gamma \sim e^{-2\trm S} = e^{2\pi\big[(M -\omega)^2 +(M -\omega)\sqrt{(M -\omega)^2 -Q^2}
 -M^2 -M\sqrt{M^2 -Q^2}\big]} = e^{\Delta S_{BH}} \, ,
\ee
which is same one as that obtained in Ref. \cite{PW}.

For another special case when $\Lambda = 0$, $Q = 0$, and $\eta = 0$, the black hole metric considered
here reduces to the Schwarzschild black hole, one can derive the event horizon of the black hole before
and after a particle with energy $\omega$ is emitted, namely, $r_i = 2M $ and $r_f = 2(M -\omega)$.
According to Eq. (\ref{TA1}), the tunneling rate at the event horizon will be reduced to
\be
\Gamma \sim e^{-2\trm S} = e^{-8\pi(M -\omega/2)} = e^{\Delta S_{BH}} \, ,
\ee
which coincides with Parikh's result in the Schwarzschild black hole case.

In addition, our discussions made here can be directly extended to the anti-de Sitter case \cite{AdSM}
by changing the sign of the cosmological constant to a negative one, and also can be easily generalized
to higher dimensional spherically symmetric black holes case.

\section*{Acknowledgments}

S.-Q.Wu was supported by a starting fund from Central China Normal University and by the Natural Science
Foundation of China.

\end{document}